\documentclass[sigconf]{acmart}
\renewcommand\footnotetextcopyrightpermission[1]{} 
\usepackage{xcolor}
\usepackage{multirow}
\usepackage{siunitx}
\usepackage{soul}

\begin{document}

\title{GitEvolve: Predicting the Evolution of GitHub Repositories}

 \author{Honglu Zhou}
 \authornote{Both authors contributed equally to this research.}
 \author{Hareesh Ravi}
 \authornotemark[1]
 \email{hz289, hr268@cs.rutgers.edu}
 \affiliation{
   \institution{Rutgers University, NJ, USA}
 }

 \author{Carlos M.\ Muniz}
 \email{cmm609@cs.rutgers.edu}
 \affiliation{\institution{Rutgers University, NJ, USA}}

 \author{Vahid Azizi}
 \email{va190@cs.rutgers.edu}
 \affiliation{\institution{Rutgers University, NJ, USA}}

\author{Linda Ness}
\email{nesslinda@gmail.com}
\affiliation{\institution{Rutgers University, NJ, USA}}
\author{Gerard de Melo}
\email{gdm@demelo.org}
\affiliation{\institution{HPI, Univ.\ of Potsdam, Germany}}
\author{Mubbasir Kapadia}
\email{mk1353@cs.rutgers.edu}
\affiliation{\institution{Rutgers University, NJ, USA}}

\begin{abstract}
Software development is becoming increasingly open and collaborative with the advent of platforms such as GitHub. Given its crucial role, there is a need to better understand and model the dynamics of GitHub as a social platform. Previous work has mostly considered the dynamics of traditional social networking sites like Twitter and Facebook. We propose GitEvolve, a system to predict the evolution of GitHub repositories and the different ways by which users interact with them. To this end, we develop an end-to-end multi-task sequential deep neural network that given some seed events, simultaneously predicts which user-group is next going to interact with a given repository, what the type of the interaction is, and when it happens. To facilitate learning, we use graph based representation learning to encode relationship between repositories. We map users to groups by modelling common interests to better predict popularity and to generalize to unseen users during inference. We introduce an artificial event type to better model varying levels of activity of repositories in the dataset. The proposed multi-task architecture is generic and can be extended to model information diffusion in other social networks. In a series of experiments, we demonstrate the effectiveness of the proposed model, using multiple metrics and baselines. Qualitative analysis of the model's ability to predict popularity and forecast trends proves its applicability\footnote{To promote further research, we make our model and its variants publicly available at \url{https://github.com/hongluzhou/GitEvolve}}.

\end{abstract}



\maketitle
\fancyhead{}
\section{Introduction}

GitHub has become the most prominent collaborative software development platform. 
It resembles other social networks in that traditional social network concepts such as influence, popularity, contribution, engagement, trust etc., which have been studied extensively in the literature, apply to the social structure and temporal dynamics of key processes in GitHub \cite{borges_2016}. 
For instance, information is being shared through direct user interaction, e.g., by \emph{following} or becoming a \emph{member} of a specific project, or through indirect user--repo interactions, e.g., a user interacting by \emph{forking}, \emph{committing}, or \emph{pushing} to repositories (repos).

GitHub software is widely used by corporations and government organizations that are involved in software development. Open source components are present in about 96\% of commercial applications \cite{security_github}. 
Understanding how to simulate user behavior and repo evolution on GitHub could potentially help us identify influential users and projects and understand technological innovation and evolution. 
Due to the popularity of GitHub in open source software development, several recent studies have started to analyze specific aspects of GitHub, such as influential factors for popularity of a repo \cite{aggarwal_2014,zhu_2014} or prediction of popularity \cite{weber_2014,borges_2016}.

In this paper, we consider the more challenging task of predicting social interactions at a granular level unlike existing work \cite{borges_2016,deepcas_2017} that focuses on predicting only a single specific aspect. Specifically, we formulate the problem of simulating the evolution of a GitHub repository, i.e., given some seed events made by users to a repo and its metadata, we simulate the future sequence of events within a particular test time period. Each \emph{event} is a 3-tuple (event type, user group, time stamp). Note that, without loss of generality, this bears some resemblance to information diffusion in traditional social networks. 
\begin{figure*}[ht]
  \centering
  \includegraphics[width=\linewidth]{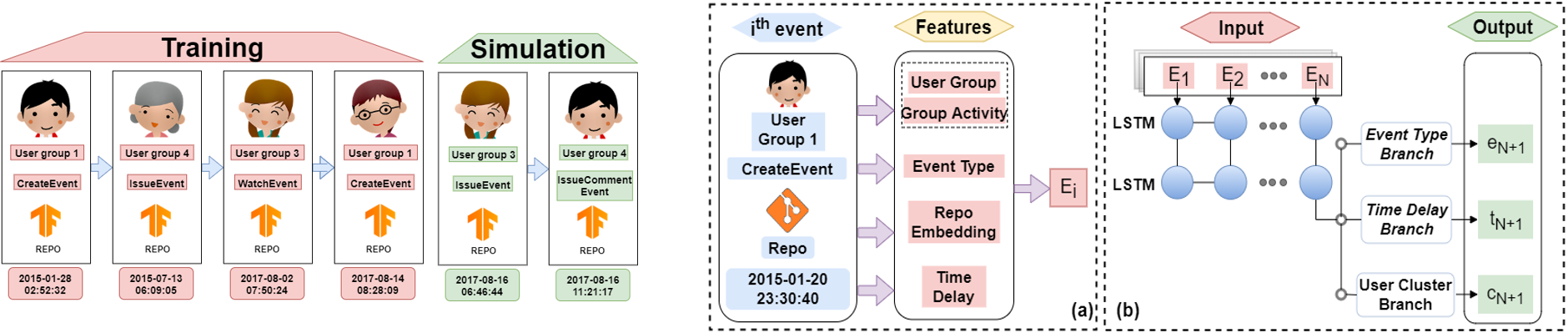}
  \caption{\emph{Left:} Example of the evolution path of a GitHub repo. Each event on this path is a 3-tuple (event type, user group, time stamp). The goal is to simulate the future evolution of this path 
  (indicated in green) 
  given some seed events seen in the past
  (indicated in red)
  . \emph{Right:} The proposed framework for fine-grained GitHub Repo Evolution Prediction. (a) shows the encoded feature for each event 
  (Sec.~\ref{sec:datarepresentation}), and 
  (b) shows the network architecture of the end-to-end multi-task sequential deep neural network ($MTS_{all}$). The designed model uses a LSTM-based encoder to encode the information of the previous $N$ events and capture their chronologically causal relationships, and employs three task-specific branches to predict the event $N+1$.}
  \label{fig:framework}
\end{figure*}
To this end, we propose an end-to-end multitask sequential deep network that recursively predicts events made to a repo. It simulates the 3-tuple characterizing each event sequentially until the designated end period of simulation is reached. The historical events made to the repo and their chronologically causal relationship are captured using a sequential deep encoder. For example, users generally \emph{fork} a project to make changes to it and potentially, but not necessarily, propagate changes back to original repo by initiating a \emph{pull request} event \cite{zhou_2019}. These interactions spanning users, repos, and events are difficult to capture with traditional sequential networks (the baseline in our experiments) or just hand-crafted features. Hence, to capture these nuances and further aid the model in simulation, graph-based automatic repo embeddings are learned that implicitly model similarities between repos via common users. Furthermore, users are grouped based on their profile and ability to produce popular content, measured as a function of the popularity of the repos they contribute to along with several other profile features. In GitHub, users that contribute to a repo by \emph{pushing} code or resolving \emph{issues} usually have write access to the repo or would have interacted with the same repo in the past. However, events such as \emph{fork} and \emph{watch} are reflective of a user's interest in the repo and do not necessarily result in a contribution to the repo. Thus, they are often one-time events. Capturing these more common, one-time events is non-trivial. We believe the user groups facilitate the learning of users that have common interests, aiding in the simulation of popularity-related events.

Our contributions are, (i) We propose a multi-task architecture for solving the fine-grained task of repo evolution prediction that simultaneously predicts three governing aspects of a GitHub event (user group, time and event type) recursively. (ii) We create user groups based on features that represent each user's potential in creating popular content. We also propose a set of features that characterize the activity of each group and show its efficacy through ablation studies. (iii) We introduce an artificial event type to better cope with long periods of inactivity or contrasting activity between training and simulation period. This leads to significant increases in simulation fidelity. (iv) We automatically learn graph-based repo embeddings that model the social structure of GitHub repos based on their description and the common interests of the user that created the repo. These embeddings help the model perform better in simulation experiments. (v) The effectiveness of the proposed technique and the importance of its various aspects are comprehensively evaluated, using an array of evaluation metrics. To the best of our knowledge, we believe this to be the first paper to propose the problem, a solution, and a comprehensive evaluation of the task of repo evolution prediction. The experimental evaluation shows the proposed model performs better than alternatives. Beyond this, our method designed to model the evolution of a GitHub repo can also be extended to model information diffusion in other social networks, and could further help with downstream tasks (e.g., popularity prediction).

\section{Related Work}
\label{sec:relatedwork}
There are multiple studies that analyze and seek to understand the dynamics of various social networks. Predictive techniques mostly tend to fall within two specific formulations: predicting the popularity of a specific entity, or predicting a specific aspect of information diffusion within a social network. 

\subsection{Popularity Prediction}
Predicting the popularity or virality of a specific entity within a social network is a fairly well-studied problem. For example, \cite{li_cikm_2013} predicts popularity of videos, while \cite{khosla_2014}, \cite{Deza_2015_CVPR} and \cite{guidotti_2019} proposed techniques to predict popularity of images in social media. 
A detailed survey on popularity prediction techniques is provided in \cite{gao_2019}.  
Twitter has also seen its fair share of popularity prediction methods for hashtags \cite{fang_2018} and tweets \cite{wang_2018}. \cite{darpaosu} proposed a technique to predict the vulnerability of a CVE (Common Vulnerabilities and Exposures) code based on tweets. Most recently, GitHub has started to receive attention with respect to popularity prediction. \cite{zhu_2014} and \cite{aggarwal_2014} study the importance of standard folders and proper documentation respectively on project popularity.
\cite{weber_2014} and \cite{borges_2016} use machine learning techniques to predict number of \emph{stars} that a repo receives. Similarly, \cite{borges_2016_icsme} studied the same from the perspective of programming languages and application domains. Important factors and dynamics of a \emph{star} event in GitHub were highlighted by means of an extensive user study by \cite{borges_2018}.

Although there has been substantial research on popularity prediction, previous work has been limited to a specific aspect of popularity and has not been dynamic in nature. Changes to popular GitHub repos occur frequently and a single aspect such as \emph{stars} may not entirely help predict vulnerabilities in the code that is exploited for cyber-attacks \cite{darpaosu}. The number of \emph{stars} is a dynamic measure depending on a multitude of factors, including consistency in issue resolution, competing repositories, etc. Fine-grained simulation of events between user groups and repository implicitly accounts for the dynamic nature of popularity. 

\subsection{Information Diffusion}

Information diffusion is the process of an information entity spreading within or across social networks by users that adopt and share this information. Examples of such entities include hashtags in Twitter, posts on Facebook, etc. 
The dynamics of the \emph{cascade} of events that unfold, resulting in the spread of the particular information entity, is well studied in the literature. For example, \cite{cheng_2014} and \cite{cheng_2016} proposed a technique to predict number of \emph{reshares} a post will get and bursts of activity respectively.  
\cite{deepcas_2017} proposed a deep learning method to learn a representation for a cascade and use it to predict the size of cascades in the future. Techniques like \cite{rong_2016,xu_2018} try to capture the underlying model of diffusion that causes the spread, while \cite{jung_2018} studied the spread of information from Twitter to online media such as news articles. 

The problem of repo evolution prediction proposed in this paper is similar to the problem of information diffusion. However, instead of predicting a specific aspect of the diffusion such as the underlying model or the size and the factors of diffusion, we intend to predict all aspects of the diffusion.
The closest techniques to ours are \cite{wenjian_2018} and \cite{blythe_2019}. \cite{wenjian_2018} proposed to predict the entire diffusion path of users that share a particular image using Tree-LSTMs (Long Short Term Memory). 
However, their model is not designed to predict how fast the spread happens. According to \cite{cheng_2016}, a recurrence of bursts of activity is a widespread phenomenon in cascades. In order to fully understand the virality and popularity of posted content, it is vital that the temporal dynamics of the diffusion path be well understood. \cite{blythe_2019} discuss their methods as part of a challenge of large-scale simulation of the evolution of the GitHub network. The focus is on large-scale simulation while the important factors that govern the evolution of GitHub are not well-examined.

\section{Proposed Method}
\label{sec:proposedmethod}

\subsection{Problem Statement}
\label{sec:problemstatement}
Let $ \{ R^1, R^2, ... , R^a \} $ be the set of repositories and $ \{ U_1, U_2, ... , U_b \} $ the set of users in the data under consideration, where $a$ and $b$ are the total number of repos and users, respectively. Let the sequence of events made to any repo $R^k$ during the training period be $ \{ E_1^k, E_2^k, ... , E_n^k \}$, where $E_i^k$ is characterized by the 3-tuple $(e_i^k, c_i^k, t_i^k)$ and denotes the $i$-th event to the $k$-th repo. Here, $e_i^k$ is the type of event made to repo $R_k$ at time $t_i^k$ by user $U_i^k$ in the group $c_i^k$. Then, the repo evolution prediction task is to predict the sequence of events $ \{ E_{n+1}^k, E_{n+2}^k, ... , E_{n+m}^k \}$ that happen during the simulation period. Since most repos do not have an actual event that denotes an $END$ (last event), we simply elect to stop the prediction when the time stamp $t_{n+m}^k$ of an event $E_{n+m}^k$ goes beyond the designated simulation period. An example that illustrates the entire problem statement is given in Fig.~\ref{fig:framework}. In order to simplify the notation, the superscript $k$ that denotes the repo is omitted in the rest of the paper. All definitions hereafter refer to a single repo and apply equally for all other repos in the data. The 3-tuple that characterizes each event $E_i$ is then $(e_i, c_i, t_i)$, where $i \in \{1, 2, ..., n\}$.

\subsection{Feature Representation}
\label{sec:datarepresentation}
We describe the types of information and their representation that is given as input to the model for simulation. 

\noindent \textbf{Event Type History.}
Event type $e_i$ is encoded as $onehot(e_i)$. The event types considered are \emph{Create}, \emph{Delete}, \emph{Fork}, \emph{Issues}, \emph{Issue Comment}, \emph{Pull Request}, \emph{Pull Request Review Comment}, \emph{Push}, \emph{Commit Comment} and \emph{Watch} \cite{githubapi}.
We add the \emph{<SOC>} event label to indicate start of an event chain. We introduce another event called \emph{NoEventForOneMonth} for every month of inactivity, to model unseen changes in activity and long periods of inactivity between recurring events. This gives a total of $12$ distinct event types.   

\noindent \textbf{Time Delay History.}
We use time delay to encode time difference between events. Specifically, time delay of event $e_i$ is defined as the number of hours since the last event and calculated as $(t_i - t_{i-1})/3600$, where $t_i$ are UTC seconds since January 1, 1970. This has a wide range and is hence $\log$-normalized as $10\log_{10}(t_i + 1)$ before being provided as input to the model. 
\begin{table}[h]
  \centering
  \small
  \begin{tabular}{@{}p{.25\linewidth}lp{.25\linewidth}lp{.25\linewidth}@{}}\toprule
      \multicolumn{1}{c}{USER PROFILE} &&\multicolumn{1}{c}{REPO PROFILE} &&\multicolumn{1}{c}{USER ACTIVITY} \\\cmidrule{1-1}\cmidrule{3-3}\cmidrule{5-5}
      One-hot encoding of user type (2) && \multirow{ 4}{\linewidth}{Most common programming language (1) }&&\multirow{ 4}{\linewidth}{Mean time (in hours) between two events (1 + 1)}\\\cmidrule{1-1}
      Country, GitHub Impact \cite{ghimpact} (2)&&&& \\ \cmidrule{1-1}\cmidrule{3-3}\cmidrule{5-5}
  \# Followers, \# Followees (2) &&\multirow{ 4}{\linewidth}{One-hot of type of user that created the repo (2)} &&\multirow{ 4}{\linewidth}{Variance of time (in hours) between two events (1 + 1)}\\\cmidrule{1-1}
     \# Repos created by the user (1)&&&& \\\cmidrule{1-1}\cmidrule{3-3}\cmidrule{5-5}
      \# forks, \# watches for created repos (2)&&Average word vector of repo description (150)&&Total count per event type (12 + 12)\\\bottomrule
   \end{tabular}
   \vspace{1em}
  \caption{User and Repo specific features. \lq{}\#\rq{} represents \lq{}number of\rq{} and \lq{}(n)\rq{} represents the feature length. All features are extracted once over the entire training period. User type could be individual or organization. User activity has one vector for specific repo being simulated and another over all repos that user has interacted with. 
  }
  \label{tab:features}
\end{table}

\noindent \textbf{User Grouping and Encoding.}
The user information associated with each event is encoded as follows. For each user in the data, their profile features are extracted from the user profile information as given in Table \ref{tab:features}. The k-Means \cite{kmeans} algorithm is invoked to induce a set of user groups from these features. The features used to group the users are chosen so as to characterize their popularity since users who are popular often tend to be those who actively maintain and contribute to repos that gather a lot of attention (cf. Table \ref{tab:features}). 
Once users are mapped to groups, two types of features are calculated. A one-hot vector that denotes which group the current user belongs to, is denoted as $onehot(c_i)$. User activity features given in Table \ref{tab:features}
are averaged over all users within a group to obtain the user group activity features, depicted as $activity(c_i)$. User group activity features are log-normalized the same way as time delay.

\noindent \textbf{Graph-based Repository Embedding.}
Repo evolution prediction is formulated from the perspective of characterizing repos and their behaviors. In order to be able to predict a repo's evolution, one has to not only encode features of that specific repo but also how it relates to any other repo in the data. Instead of directly using the repo's profile features or hand-crafting various features, we rely on a graph-based approach to extract repo features.
Thus, the model is better able to predict events for unpopular repos or for repos with fewer events in the training period. To obtain a meaningful representation of the characteristics of a repo, we induce a graph in which repos serve as nodes and there is an edge between two repos if both were created by a common user. As node attributes, repo profile features (Table \ref{tab:features}) that are part of the data are used. This attributed graph then serves as input to a graph-convolutional node representation learning process \cite{graphsage}. This yields 256-dimensional representations, which we denote as $f(R)$, where $f(\cdot)$ maps each repo to its corresponding representation.

\subsection{Multi-Task Sequential Model}
\label{sec:proposedmodel}

The entire framework of the proposed system is shown in Fig.~\ref{fig:framework}. At each time step, the model takes in the previous $N$ events and generates three outputs characterizing the next event (i.e., event type, time delay, and user group) simultaneously. The \textit{Event Type} branch outputs a vector that provides the probability distribution of event $e_{i+1}$ being each of the $12$ event types. The \textit{Time Delay} branch provides the log-normalized time delay value $t_{i+1}$. The \textit{User Group} branch emits a vector that gives the probability that $c_{i+1}$ is any of the $C$ user groups. The predicted vectors are then processed and added as input to predict the events for subsequent time steps.

The first half of the multi-task network consists of a two-layered LSTM network that sequentially encodes the $N$ input events. This ensures that the pattern of events, the interaction between sequence of users and the frequency with which events are made to the repo are modeled. The final encoded vector then serves as input to the second half of the network. 
\begin{table*}[ht]
  \centering
  \begin{tabular}{@{}cp{0.1\linewidth}llllllllllll@{}}\toprule
    &\textbf{Model} & \multicolumn{5}{c}{\textbf{Event-Type}} & \textbf{Time-Delay} &\multicolumn{5}{c}{\textbf{User-Group}}& \textbf{\#Events} \\\cmidrule(lr){3-7}\cmidrule(lr){8-8}\cmidrule(lr){9-13}\cmidrule(lr){14-14}
    &&mAP&$\text{BLEU}_{1}$&$\text{BLEU}_{2}$&$\text{BLEU}_{3}$&$\text{BLEU}_{4}$&DTW &mAP&$\text{BLEU}_{1}$&$\text{BLEU}_{2}$&$\text{BLEU}_{3}$&$\text{BLEU}_{4}$&MAE\\\midrule
    &Random&0.04&0.03&0.16&0.00&0.00&\num{14.6}&0.42&0.01&0.04&0.00&0.00&22.80\\\cmidrule{2-14}
    &Previous&0.41&0.03&0.16&0.07&0.07&\num{4637}&0.40&0.01&0.04&0.06&0.06&\num{14469}\\\cmidrule{2-14}
    &NoEvent&0.40&0.03&0.16&0.00&0.00&\textbf{0.07*}&0.01&0.40&0.04&0.00&0.00&\textbf{2.10*}\\\cmidrule{2-14}
    &Baseline&0.04&0.03&0.16&0.15&0.11&30.0&0.01&0.01&0.04&0.05&0.06&161.8\\\cmidrule{2-14}
    &$MTS_{all}$-f(R)&\textbf{0.47*}&\textbf{0.85*}&\textbf{0.17*}&0.16&0.13&0.70&0.45&0.83&0.06&0.08&0.09&7.10\\\cmidrule{2-14}
    &$MTS_{all}$-a($c_i$)&\textbf{0.47*}&0.84&\textbf{0.17*}&0.17&0.15&1.20&0.45&0.82&0.06&0.08&0.09&10.4\\\cmidrule{2-14}
    &$MTS_{all}$/i(R)&0.45&0.84&0.13&0.14&0.13&1.33&0.45&\textbf{0.85*}&0.06&0.09&0.10&9.50\\\cmidrule{2-14}
    &$MTS_{all}$/p(R)&0.46&\textbf{0.85*}&0.15&0.16&0.14&0.62&0.45&\textbf{0.85}&0.07&0.08&0.09&8.50\\\cmidrule{2-14}
    &$MTS_{all-11}$&0.04&0.04&0.16&0.16&0.14&10.9&0.01&0.02&0.08&0.09&0.10&260.6\\\cmidrule{2-14}
    &$MTS_{all}$&0.46&\textbf{0.85*}&0.16&\textbf{0.18*}&\textbf{0.16*}&\textbf{0.28*}&\textbf{0.46*}&\textbf{0.85*}&\textbf{0.08*}&\textbf{0.11*}&\textbf{0.12*}&\textbf{5.0*}\\\bottomrule
 \end{tabular}
 \vspace{1em}
  \caption{Results for Simulation of Sequence of Events experiment. Values in bold with \lq{}*\rq{} highlight the best-performing model in that class of metric across all models. $MTS_{all}$ is the proposed model. Time-Delay DTW values have been divided by $10^3$ for brevity. Note that $\text{BLEU}_{3}$ and $\text{BLEU}_{4}$ are the more reasonable as well as challenging metrics compared to $\text{BLEU}_{1}$ and $\text{BLEU}_{2}$. $MTS_{all}$ gives the best result in $10$ out of $12$ metrics. The proposed model outperforms all other models, as it is the only one shows consistently good performance across all \textit{three} tasks.}
  \label{tab:simulation}
\end{table*}
The second half has three branches, viz., event type, time delay, and user group branches. Each branch has two fully-connected layers with the non-linear ReLU activation, along with dropout layers \cite{dropout} between the dense layers, to avoid overfitting. These branches ensure that the characteristics of each of the tasks are modelled independently while the first half combines information from all the tasks to model their dependencies. The total loss for the first part of the network is given as, 
\begin{align}
    loss &= w_{e}  \bigg(\sum_{k=1}^{e_N}-\frac{1}{bs}\sum_{b=1}^{bs}t_{k,b}\log(p_{k,b}) + (1 - t_{k,b})\log(1 - p_{k,b})\bigg)\nonumber\\
    & + w_{t}\bigg(-\frac{1}{bs}\sum_{b=1}^{bs}\log(\cosh(p_b - t_b))\bigg)\label{eq1}\\
    & + w_{c} \bigg(\sum_{k=1}^{C}-\frac{1}{bs}\sum_{b=1}^{bs}t_{k,b}\log(p_{k,b}) + (1 - t_{k,b})\log(1 - p_{k,b})\bigg) \nonumber 
\end{align}

\noindent where $w_{e}$, $w_{t}$ and $w_{c}$ represent the weights for event type, time delay and user group losses respectively and are all chosen as $1$ to weight the importance of each of the tasks equally. This also gave the best results empirically. Instead of the usual categorical loss for classification, we opt for the multinomial binary cross-entropy loss for event type and user group tasks. This ensures that the error does not depend only on the probability associated with the true class at all instants. We again treat all the classes equally irrespective of the class imbalance. This way, popular events and repos remain popular and are not balanced with unpopular events and repos.
Here, $k$ iterates over classes, and $b$ over samples in the batch. $e_N$ is the number of event type classes, while $C$ is the number of user groups. $p_{k,b}$ is the probability that the $b^{th}$ sample corresponds to class $k$, while $t_{k,b} \in \{0, 1\}$ depending on whether the sample belongs to class $k$ or not. Similarly, $p_b$ and $t_b$ correspond to the predicted value and the true value, respectively. For time delay values, we rely on the log-cosh function (Eq.~\ref{eq1}), as it is not affected by the occasional wrong prediction, but still penalizes incorrect high predictions. This is chosen over the more traditional MSE or MAE errors to handle the variance in the data. 

\section{Experimental Setup}
\label{sec:expsetup}
\subsection{Dataset}
To motivate the need for fine-grained simulation, we focus on GitHub repositories that are related to CVE \cite{cve}. 
The gharchive service \cite{gharchive} identifies repos that have CVE codes mentioned in their profile description, from which our dataset includes 700 randomly chosen CVE IDs. This set was filtered to maintain only repos with events in the period from January 1, 2015 to August 31, 2017. This yields a total of $1,377$ repos and $49,000$ users that have made at least one event during that period. For all the users and repos in the data, profile data was obtained from ghtorrent \cite{ghtorrent}.
This contributed to the profile features enumerated in Table \ref{tab:features}. 

We observe that the variance of data in general is high, adding to the complexity of the problem. For example, about 74\% of users make only 1 event, whereas just 2\% of users are responsible for about 90\% of all events. Owing to these imbalances, we opted to predict user groups rather than each individual user. Also, the evolution path length (or the number of events) of the most popular repo is $27,244$, whereas 42\% of the repos have only 2--10 events and 25\% of the repos have only 1 event. On top of this, the distribution of the number of events for each event type is also imbalanced where \emph{Watch} events make up for 30\% of the events. Whereas, \emph{Delete}, \emph{Commit Comment} and \emph{Pull Request Review Comment} events make up for only 5\% of total events put together. It is a common practice to constrain the data \cite{wenjian_2018} based on activity or popularity, to simplify the modeling. However, we opt to perform little to no filtering. The only constraint is to retain the event types given in Sec.~\ref{sec:proposedmethod}, which are more common, while ignoring other events such as the Gollum Event \cite{githubapi}, as they are not present for the vast majority of repos. The training period is from January 1, 2015 to July 31, 2017. Data for validation came from events that happen within August 1--15, 2017, and the simulation period is set as August 16--31, 2017. Simulation is performed only for repos that have at least one event during the training period.

\subsection{Implementation Details}
The hidden sizes of the first two LSTM layers are set as $250$ and $150$, respectively. The following two dense layers for all the branches are set to be $128$ and $64$, respectively. All dropout layers have a dropout value of $0.5$. We train the model for $150$ epochs with the Adam \cite{adam} optimizer, and choose the epoch that has the lowest validation loss for simulation. The number of input events $N$ at each time step and the batch size are set as $20$ and $256$, chosen empirically using validation data, to get the best result from sets $\{10, 20, ..., 200\}$ and $\{32, 64, 128, 256, 512\}$, respectively.  The number of user groups $C$ was empirically chosen to be $100$ after elbow curve \cite{elbow}
and silhouette \cite{silhouette} analyses. One extra group of users $c_{C+1}$ is added as part of events with $No Event For One Month$ event type, giving rise to a total of $101$ groups. Group activity features are $28$-d (cf. Table \ref{tab:features}). The embedding dimension of repository is $256$. Different values $\{64, 128, 256, 512\}$ were tried using the validation data before choosing the best-performing one. The total length of input features to the model is $12$ + $1$ + $101$ + $28$ + $256$ $=$ $398$. 

\section{Evaluation}
\label{sec:evaluation}
\subsection{Baselines}
Due to the novelty of the problem, we could not find any state-of-the-art work that addressed the proposed problem for comparison. Existing approaches that analyze GitHub \cite{borges_2016} remain inapplicable for the majority (if not all) of the metrics reported here, which require explicit fine-grained prediction. Hence, it is not possible to compare the proposed technique with these approaches. However, in order to better understand the effect of the proposed set of features and the experimental setup, we performed comprehensive ablation studies and compare with intuitive baselines. The following models are evaluated to demonstrate the performance of the proposed model.

$\mathbf{Random}$ is based on predicting the Event-Type and User-Cluster from set of possible event types and user clusters. The Time-Delay is predicted randomly from the range defined by minimum and maximum time delay values from events made previously to the repo. This gives a clearer reference point to assess the performance of the proposed model.

$\mathbf{Previous}$ repeats the previous action until the end of the testing period.

$\mathbf{NoEvent}$ repeats the \emph{NoEventForOneMonth} label until the end of the testing period.

$\mathbf{Baseline}$ is a simple Multi-Task Sequential LSTM model without \emph{NoEventForOneMonth}, user group activity or repo embeddings.

$\mathbf{MTS_{all}}$ is the proposed model with user group activity features, repo embeddings, and additional \emph{NoEventForOneMonth}.

$\mathbf{MTS_{all}-f(R)}$ does not have repo embedding vector $f(R)$.

$\mathbf{MTS_{all}-a(c_i)}$ has no user group activity features $a(c_i)$.

$\mathbf{MTS_{all}/i(R)}$ is trained  by replacing the repo embedding vector $f(R)$ with a one-dimensional log-normalized repo index $i(R)$.

$\mathbf{MTS_{all}/P(R)}$ replaces the repo embedding vector $f(R)$ with the repo profile features $P(R)$ outlined in Table \ref{tab:features}.

$\mathbf{MTS_{all-11}}$ has only $11$ event types. In comparison to the proposed model, it lacks the extra \emph{NoEventForOneMonth} event type.
\subsection{Evaluation Type}
\noindent \textbf{Simulation of Sequence of Events.}
For simulation, the model sees the previous $N$ events ($N = 20$, cf.\ Section \ref{sec:expsetup})
before the start of simulation period and recursively predicts. After each prediction, the predicted values are concatenated with the previous $N-1$ events to make up the $N$ input events for the next event's prediction. The evaluation proceeds for all $1,377$ repos in the dataset. If a repo does not have any event during the simulation period, the model is expected to predict the lack of events. For the sake of evaluation, repos without any event during simulation period are evaluated as having a single $NoEventInSimPeriod$ event. This is in order to have non-zero $\text{BLEU}_{1}$ and mAP values for these repos. If the model does not predict any events for this repo, the model is treated as having output $NoEventInSimPeriod$. Otherwise, whatever events the model predicts are retained. 
\begin{table*}[h]
  \centering
  \begin{tabular}{@{}cp{0.1\linewidth}lllllllllll@{}}\toprule
    &\textbf{Model} & \multicolumn{5}{c}{\textbf{Event-Type}} & \textbf{Time-Delay} &\multicolumn{5}{c}{\textbf{User-Group}}
    \\\cmidrule(lr){3-7}\cmidrule(lr){8-8}\cmidrule(lr){9-13}
    &&mAP&$\text{BLEU}_{1}$&$\text{BLEU}_{2}$&$\text{BLEU}_{3}$&$\text{BLEU}_{4}$&DTW&mAP&$\text{BLEU}_{1}$&$\text{BLEU}_{2}$&$\text{BLEU}_{3}$&$\text{BLEU}_{4}$\\\midrule
&Baseline&0.49&0.69&\textbf{0.59*}&0.43&0.42&931&0.16&0.26&0.28&0.29&\textbf{0.32*}\\\cmidrule{2-13}
    &$MTS_{all}$-f(R)&0.80&0.86&0.46&0.47&0.40&162&0.78&0.80&0.23&0.29&\textbf{0.32*}\\\cmidrule{2-13}
    &$MTS_{all}$-a($c_i$)&0.80&0.86&0.50&0.47&0.43&130&0.78&0.80&0.23&0.27&0.28\\\cmidrule{2-13}
    &$MTS_{all}$/i(R)&0.80&0.85&0.46&0.47&0.40&176&\textbf{0.78*}&0.81&0.22&0.27&0.31\\\cmidrule{2-13}
    &$MTS_{all}$/p(R)&0.81&0.86&0.47&0.47&0.41&150&\textbf{0.78*}&0.80&0.24&0.30&0.31\\\cmidrule{2-13}
    &$MTS_{all-11}$&0.49&0.68&0.57&\textbf{0.48*}&\textbf{0.47*}&898&0.15&0.28&0.23&0.30&\textbf{0.32*}\\\cmidrule{2-13}
    &$MTS_{all}$&\textbf{0.82*}&\textbf{0.87*}&0.48&\textbf{0.48*}&0.41&\textbf{129*}&\textbf{0.78*}&\textbf{0.82*}&\textbf{0.31*}&\textbf{0.31*}&\textbf{0.32*}\\\bottomrule
\end{tabular}
 \vspace{1em}
  \caption{Results for Prediction of Single Event experiment. Values in bold with \lq{}*\rq{} denote the best-performing in that class of metric across all models. The proposed model $MTS_{all}$ gives the best performance in $9$ out of $11$ metrics.}
  \label{tab:prediction}
\end{table*}

\noindent \textbf{Prediction of Single Event.}
Following the experimental setup of \cite{wenjian_2018}, we evaluate the ability of the model to predict the \lq{}next\rq{} event and all its aspects correctly, given the ground truth sequence of previous events. At each time step, the previous $N$ ground truth events are given to predict the next event in the simulation period. This process continues until there are no further ground truth events available. The predicted events are then compared against the ground truth. 
Evaluation occurs only for repos that have at least one event in the testing period, as there will be no ground truth to give afterwards. This brings the number of repos to $847$ for $\text{BLEU}_{1}$, $145$ repos for $\text{BLEU}_{2}$, $97$ for $\text{BLEU}_{3}$ and $82$ for $\text{BLEU}_{4}$. 

\subsection{Evaluation Metrics}
To quantitatively and comprehensively evaluate the simulated sequence of events by comparing it with the ground truth, we rely on the following array of metrics.

\noindent \textbf{mAP.} 
mean Average Precision is used to evaluate the precision of event type and user group prediction tasks following the usage in \cite{wenjian_2018} for prediction experiment and \cite{maptweet} for simulation experiment. The range of the metric is between 0 and 1, 1 being the best.

\noindent \textbf{$\text{BLEU}_{k}$.} 
Traditional generation tasks in natural language processing use the \emph{$\text{BLEU}_{k}$} \cite{bleu} metric to measure how close a generated caption is to a ground truth caption. 
We evaluate the event type and user group sequences with the \emph{$\text{BLEU}_{k}$} metric,
where $k \in {1, 2, 3, 4}$ applies only on chains that have at least \lq{}$k$\rq{} events. The range of the metric is between 0 and 1 with 1 being the best. 

\noindent \textbf{DTW.} 
Dynamic Time Warping \cite{dtw} is used to evaluate two temporal sequences and we use this to evaluate the time-delay prediction task. The lower the DTW value, the better.

\noindent \textbf{MAE.} 
We provide the Mean Absolute Error between the total number of events predicted and that in the ground truth for all the repos for the simulation experiment. For the prediction experiment, the number of events predicted will be same as in the ground truth.
\section{Results}

The results for the prediction of single event and simulation of sequence of events experiments are given in Table \ref{tab:prediction} and Table \ref{tab:simulation} respectively. It is to be noted that $\text{BLEU}_{1}$ is significantly higher than $\text{BLEU}_{k}$, when $k > 1$, especially for the simulation experiment. This is because the model is able to predict inactivity accurately. Since all repos that do not have any event during the simulation period are evaluated as part of $\text{BLEU}_{1}$ and mAP, these scores are considerably higher. 

Overall, although there are several cases where other model variants perform slightly better in the prediction experiment, the proposed $MTS_{all}$ model shows consistent performance across all metrics and tasks in both simulation and prediction experiments. The proposed model obtains the best results with regard to 10 out of 12 metrics in the simulation experiment and for 9 out of 11 metrics in the prediction experiment. Most importantly, our model is consistent across tasks, whereas other models perform well on either just one task at a time or under-perform across all the tasks, as observed in Table~\ref{tab:simulation}.
\begin{figure*}[h]
  \centering
  \includegraphics[width=0.7\linewidth]{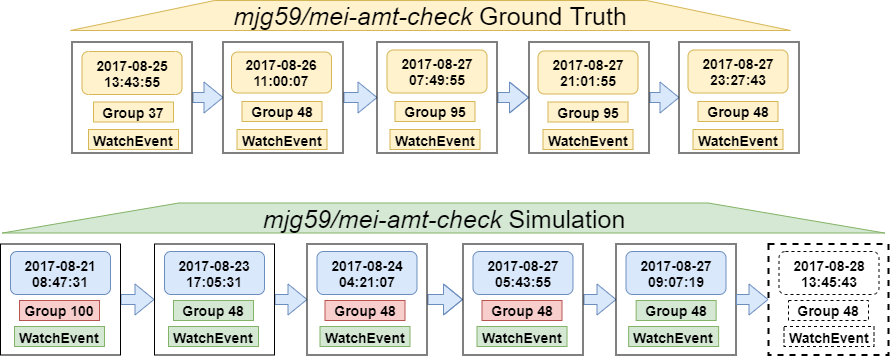}
   \vspace{0pt}
 \caption{GT and simulated sequence of events for repo \emph{mjg59/mei-amt-check}. The top (\emph{yellow}) shows the ground truth, \emph{red} indicates wrong predictions, \emph{green} indicates correct predictions. Dotted boxes denote events not part of the ground truth.}
 \label{fig:result}
\end{figure*}
\begin{figure}[h]
  \centering
  \includegraphics[width=0.7\linewidth]{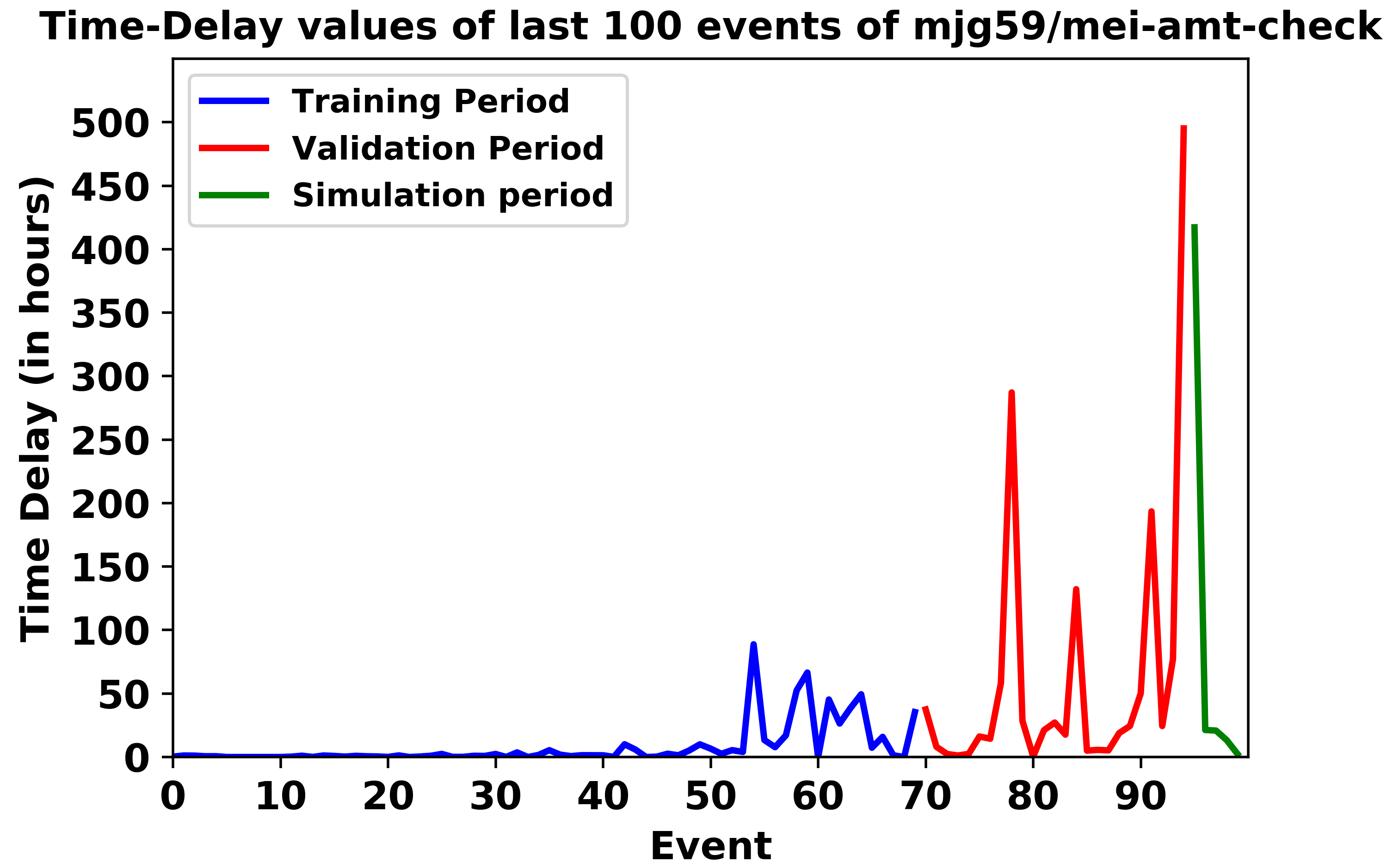}
  \vspace{0.5em}
  \includegraphics[width=0.7\linewidth]{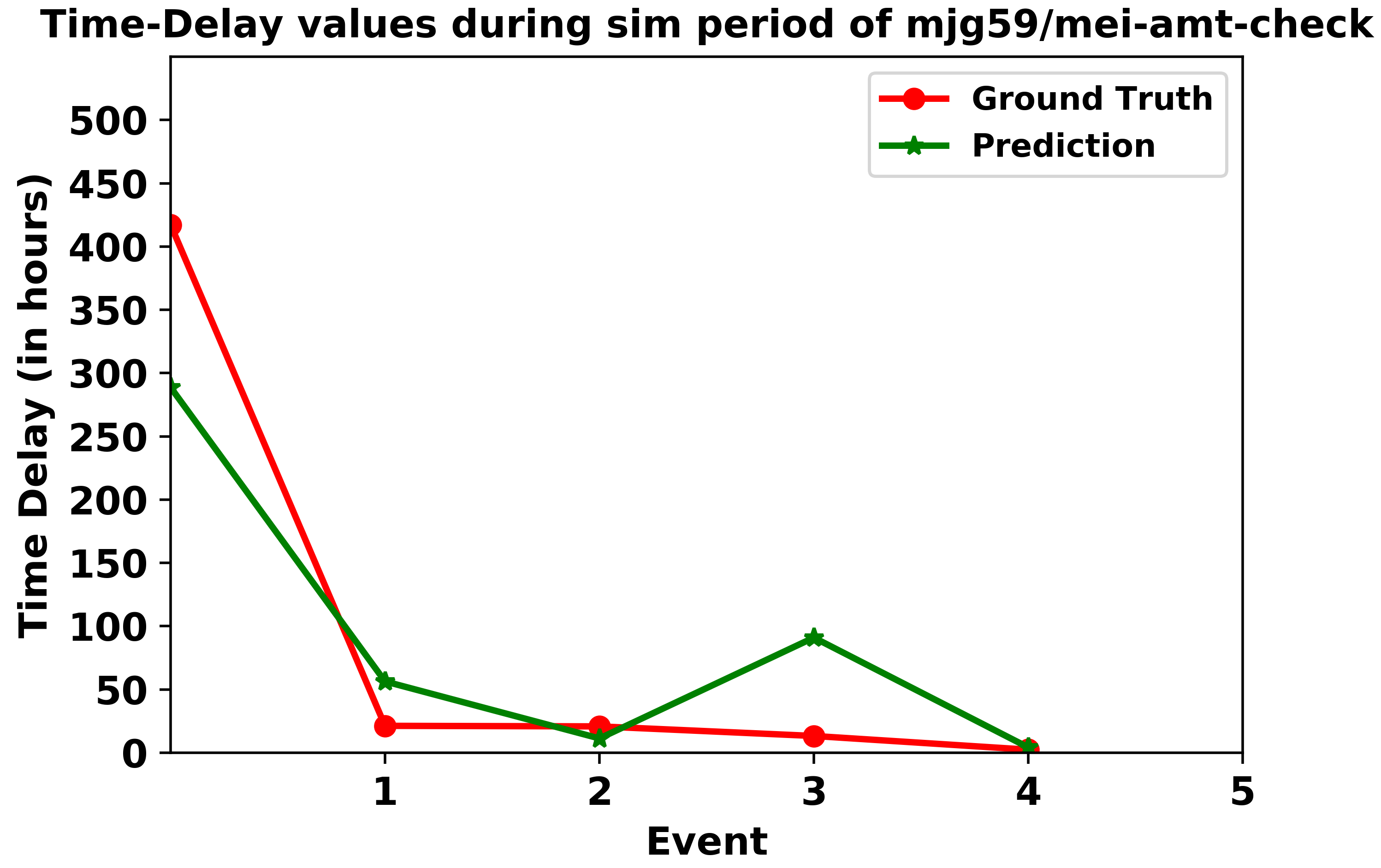}
 \caption{\emph{Top:} Time delay values of last 100 events of repo \emph{mjg59/mei-amt-check} show contrasting activity between the training and simulation periods indicating the complexity of the prediction problem.
 \emph{Bottom:} shows the ground truth v.s. the simulated time delay values for the same repository given the complexity of the prediction problem.}
  \label{fig:timedelay}
  \vspace{-0.1in}
\end{figure}

An example simulation result along with the ground truth is provided in Fig.~\ref{fig:result} for a randomly chosen repo \emph{mjg59/mei-amt-check}. This repo mentions the CVE ID `CVE-2017-5689' in its description. Fig.~\ref{fig:timedelay} (left) shows the time delay values of the last 100 events for this repo. Contrasting behavior in activity between training, validation and simulation period is clearly observed. 
Fig.~\ref{fig:result} shows the actual predictions. We are able to predict event types perfectly. Predicting the exact time stamp is an extremely hard problem. Although our predictions are off by a day or two, we are successfully able to predict the general pattern in activity. we provide a plot of the time delay values for the GT versus the simulation for this repo for better visualization in Fig.~\ref{fig:timedelay} (the extra predicted event is not plotted). The user group prediction however, is far from perfect owing to high variance and large number of user groups.

\section{Discussion}

\noindent \textbf{Are group activity features important?}
Removing user group activity features ($\mathbf{MTS_{all} - activity(c_i)}$) does not hamper the performance in the prediction experiment as much as one would expect. However, in the simulation experiment, the average DTW score for the time delay is significantly worse compared to other models with the activity features. This is because the type of events made to a repo depends on historic user activity and the popularity of the repo as much as it does on historic events. Specifically, event types such as \emph{Watch} and \emph{Fork} happen more randomly compared to say \emph{PushEvent}, for which the possible set of users usually consists of the repo's core contributors. There are users that perform just these \emph{Watch} and \emph{Fork} actions without making contributions to any of the repos in the data. These complexities make group activity features necessary, to encode the frequency with which events of multiple types happen. Also, it can be noted that, in general, the event type task obtains better scores compared to the user group task across all models. This can be attributed to the presence of users with a high variance in activity across the dataset.

\noindent \textbf{Is repo embedding learning necessary?}
Experiments that remove repo embedding features or replace them (cf.\ Table \ref{tab:simulation}) exhibit lower performance compared to the proposed model, highlighting the importance of learning complex representations to encode repo information. 
One could explore additional graph representation learning algorithms such as \cite{graphattn} or learning the representation as part of the model itself, as proposed in \cite{deepcas_2017}. 

\noindent \textbf{Is \emph{NoEventForOneMonth} important?}
One change that yields surprisingly good results is the addition of the event type \emph{NoEventForOneMonth} as discussed earlier. This limits the maximum time delay value to be $720$ hours (1 month). The reduction in the variance, even though not significantly, helps to obtain a better-performing model. The $MTS_{all-11}$ Model without the additional \emph{NoEventForOneMonth}, performs the worst compared to other models, as can clearly be seen from the metrics. Interestingly, time delay metrics are the most affected by this change (cf.\ better values for $\text{BLEU}_{3}$, $\text{BLEU}_{4}$ for $MTS_{all-11}$ Model in Table \ref{tab:simulation}), establishing the effect of the additional event type. 
\begin{table}[h]
  \begin{tabular}{@{}ccc@{}}\toprule
    \textbf{Event Type Metric} &\multicolumn{2}{c}{\textbf{Model}}  \\\cmidrule(lr){2-3}
     &$STS_{all} - c_i$ &$STS_{all}$\\\midrule
      mAP &0.07&\textbf{0.80*}\\\midrule
      $\text{BLEU}_{1}$ &0.11&\textbf{0.85*}\\\midrule
      $\text{BLEU}_{2}$ &0.3&\textbf{0.49*}\\\midrule
      $\text{BLEU}_{3}$ &0.27&\textbf{0.45*}\\\midrule
      $\text{BLEU}_{4}$ &0.21&\textbf{0.41*}\\\bottomrule
  \end{tabular}
 \vspace{1em}
 \caption{Ablation study shows the necessity of user grouping. We train two single task model with ($STS_{all}$) and without ($STS_{all} - c_i$) user group information and perform the event type prediction experiment. The results indicate removing user group information degrades performance of the model.}
\label{fig:discussion2}
\end{table}
\begin{figure*}[ht]
  \centering
  \includegraphics[width=0.3\linewidth, height=4cm]{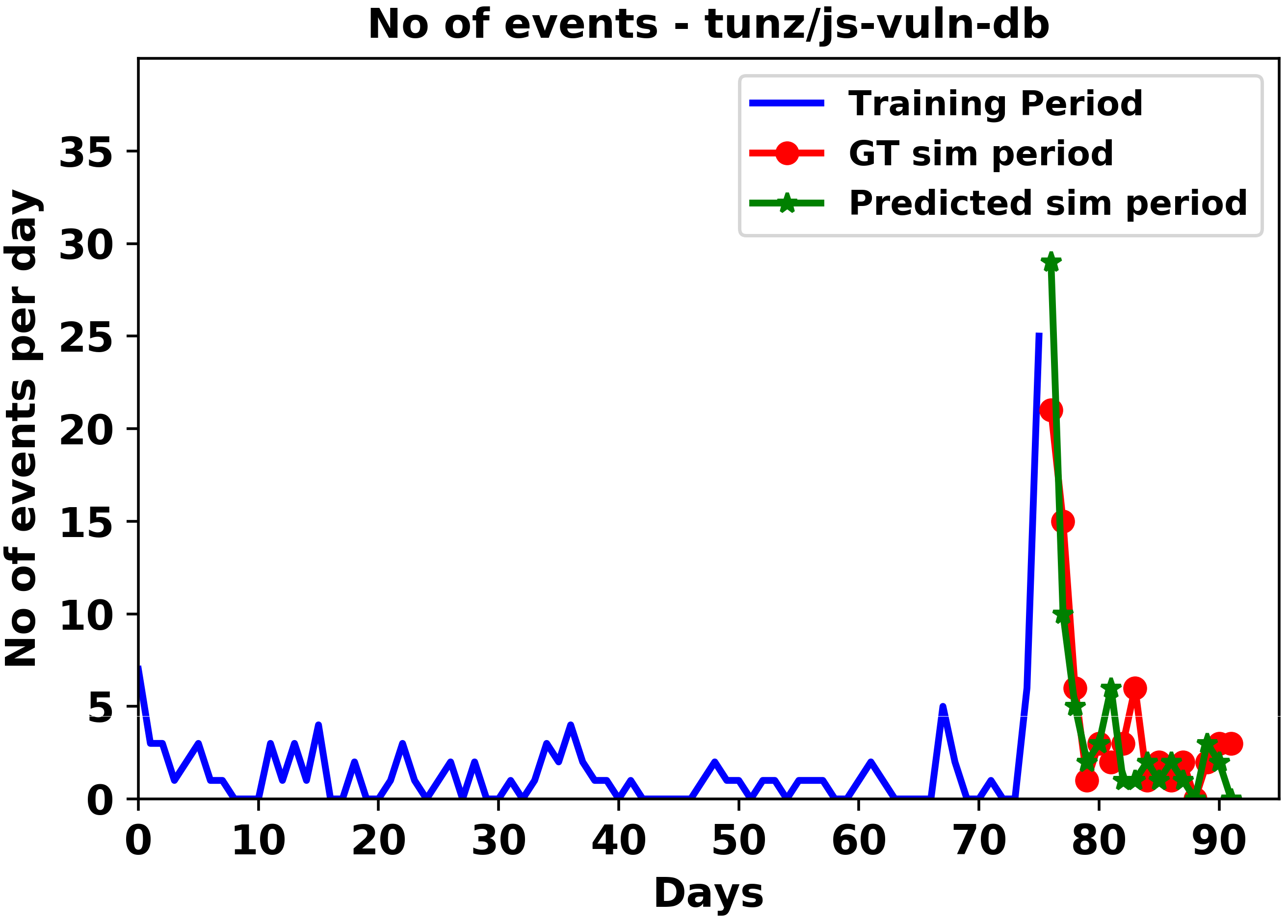}
\hspace{0.03\linewidth}
  \includegraphics[width=0.3\linewidth, height=4cm]{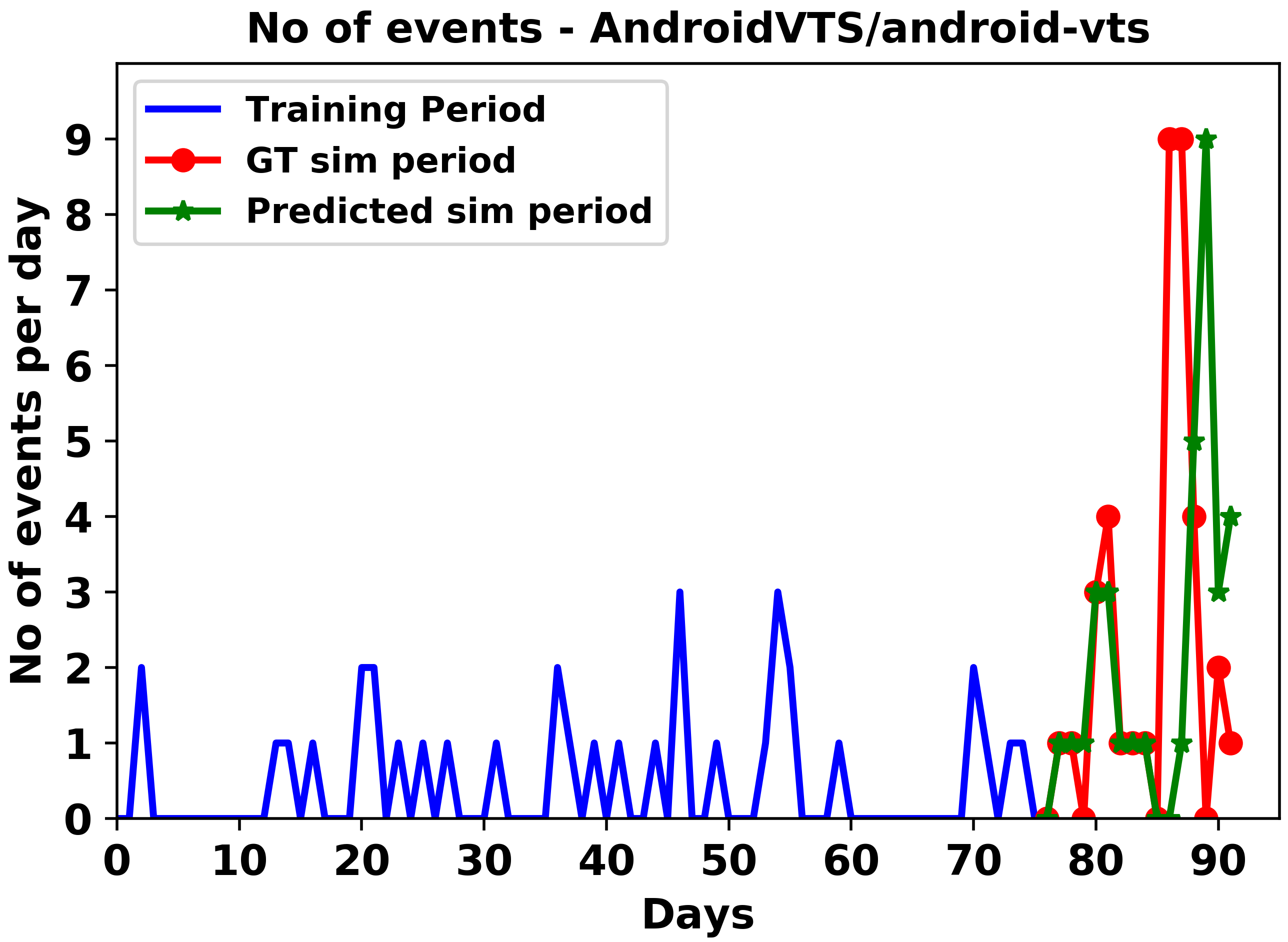}
\hspace{0.03\linewidth}
  \includegraphics[width=0.3\linewidth, height=4cm]{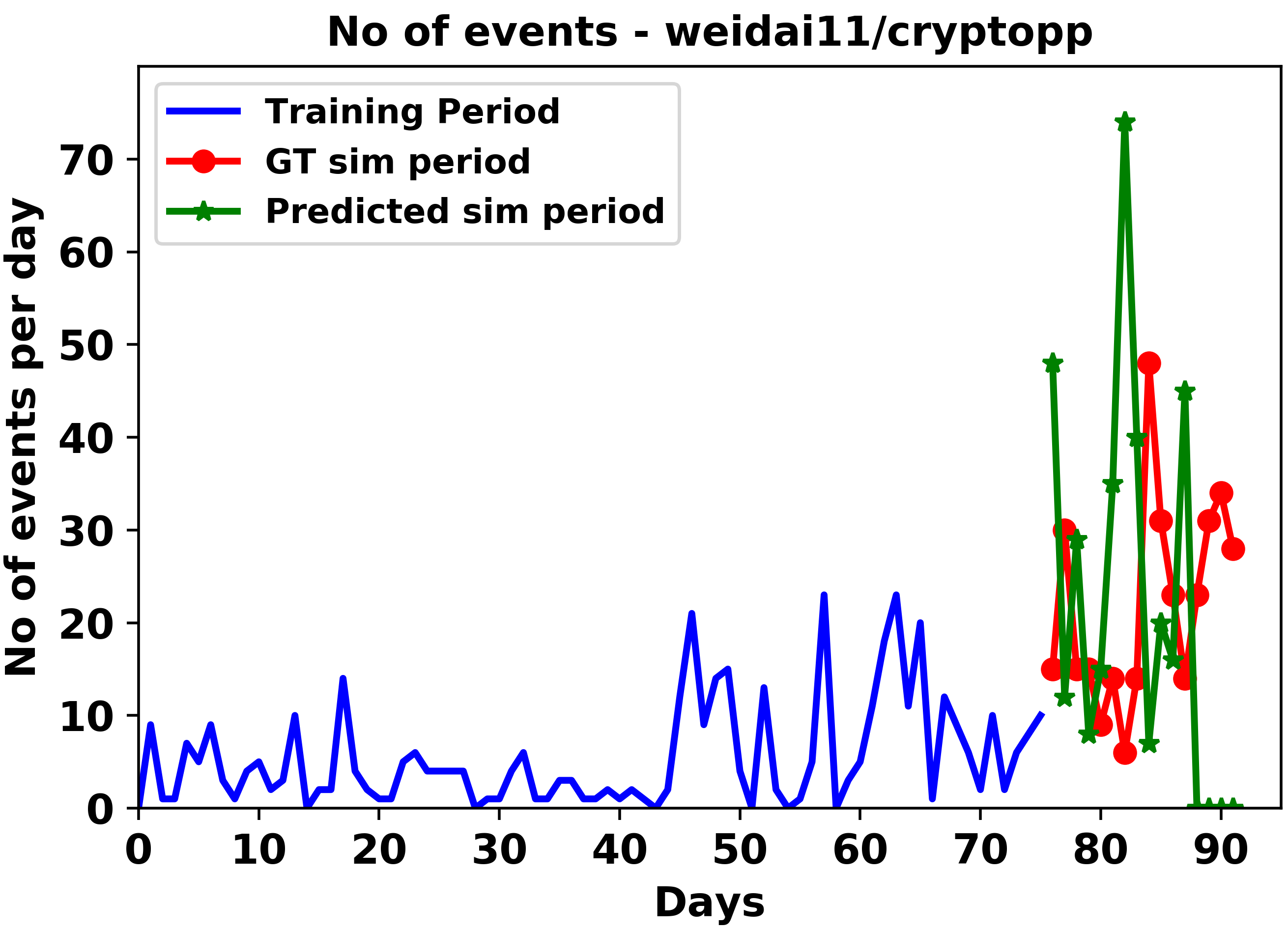}
 \caption{Ground Truth vs Predicted No of events compared with no of events during training period of three different repos.}
  \label{fig:repo2tehcnical}
\end{figure*}

\noindent \textbf{Does user grouping hinder/help the model?}
In order to further motivate the need of user grouping, we perform the prediction experiment (cf.\ Sec.~\ref{sec:evaluation}) for only event type branch, with ($STS_{all}$) and without ($STS_{all}-c_i$) user group information to analyze its effects. Results are shown in Table \ref{fig:discussion2}. To avoid not giving any user information for $STS_{all}-c_i$ model, all user activity features from Table \ref{tab:features} is given.
We can see that not having user group information hinders the performance of the model significantly. About 30\% of all events is \emph{Watch} and around 70\% of those events are made by users who have made just $1$ event in the entire dataset. Without the user-grouping, all events are classified as \emph{Watch} events implying the difficulty in capturing these frequent one-time events along with other event types. 
User grouping helps with modelling the other simultaneous tasks while enabling the model generalize to unseen users. Particularly the one-timers that keep coming up which would otherwise be harder to predict.
\begin{figure}[h]
  \includegraphics[width=0.7\linewidth, height=5cm]{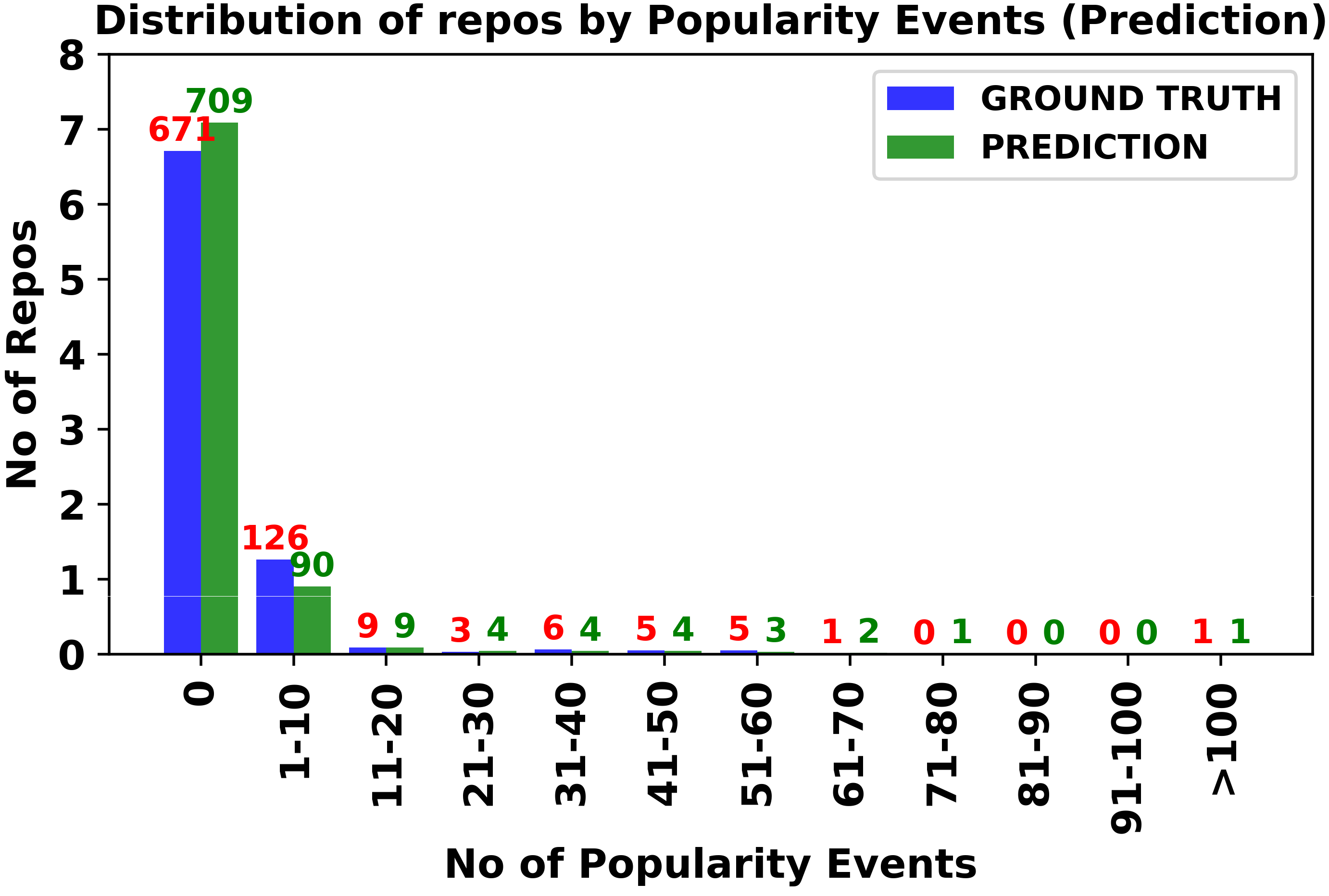}
\caption{Ground Truth vs.\ Predicted popularity events by number of repos over the entire test set.}
\label{fig:discussion1}
\end{figure}

\noindent \textbf{Can the proposed model predict popularity?}
The current model could be simplified to remove the multitask output and be trained to predict single specific aspect of popularity such as the number of \emph{Fork} events or \emph{Watch} events. However, we believe that popularity is a dynamic measure and requires a more sophisticated modelling. In Fig.~\ref{fig:discussion1}, we plot the distribution of number of repos by number of popularity-related events (\emph{Fork} and \emph{Watch} events) as predicted by the proposed fine-grained model and compare it with ground truth distribution. Although the proposed model is granular, it is able to still capture the overall popularity of different repos. We further calculated the average $MAE$ over all repositories for the prediction of the number of popularity-relatd events (\emph{Watch} and \emph{Fork}) using the proposed $MTS_{all}$ model, obtaining 1.19 for the simulation experiment and 0.85 for the prediction experiment. These low error rates indicate a high degree of accuracy in predicting the total number of popularity events by the proposed model.

\noindent \textbf{Can the model forecast technical trends?}
In order to check if the model is able to forecast technical trends, we observe the predictions of the model for three popular repos with contrasting activity between training and simulation period as shown in Fig \ref{fig:repo2tehcnical}. The proposed model is able to predict this contrasting behaviour accurately. This is because of both the graph based repo embedding features that relate different repos and the multi task architecture that models all the tasks simultaneously.

\section{Conclusion}
\label{sec:conclusion}
We present the problem of GitHub repo evolution prediction and propose a deep multi-task framework that sequentially simulates future events by simultaneously predicting a repo's event type, user group and the time stamp of the event at each time step. Due to the complex nature of the problem statement and the general difficulty of predicting the evolution of a GitHub repo, we propose a series of modeling strategies such as adding a new event type to better model periods of inactivity, graph based repo embedding to encode repo information, and user grouping based on their potential to create popular content. We demonstrate the effectiveness of this new model and features that are specifically designed for the problem at hand through a comprehensive evaluation of both per-event prediction and sequence simulation experiments. To the best of our knowledge, this is the first paper to tackle the complex problem of repo evolution prediction in a comprehensive manner. As future work, we intend to release a large dataset for this problem to promote more research. We are also investigating methods to automatically learn feature representations for users and repos simultaneously. We would like to look further into how the trained model can be used to further predict aspects such as popularity or CVE vulnerability score to prevent cyber attacks. Overall, by considering GitHub as a social network and proposing fine-granular event predictions as a novel task, we believe this work opens up promising new research avenues for the community.
\newpage
\bibliographystyle{ACM-Reference-Format}
\balance 
\bibliography{reference}

\end{document}